\documentclass[superscriptaddress,aps,pra, 10pt]{revtex4-2}

\usepackage{graphicx}
\usepackage{dcolumn}
\usepackage{bm}
\usepackage{hyperref}
\usepackage[mathlines]{lineno}
\usepackage{amsmath}
\usepackage[italicdiff]{physics}
\usepackage{xcolor}
\usepackage{multirow}

\begin{document}

\title{Spherically symmetric Earth models yield no net electron spin}

\author{N.B. Clayburn}
 \email{nclayburn@amherst.edu}
\affiliation{
 Department of Physics \& Astronomy, Amherst College, Amherst, Massachusetts 01002, USA
}
\author{A. Glassford}
\affiliation{
 Department of Physics \& Astronomy, Amherst College, Amherst, Massachusetts 01002, USA
}
\author{A. Leiker}
\affiliation{
 Department of Physics \& Astronomy, Amherst College, Amherst, Massachusetts 01002, USA
}
\author{T. Uelmen}
\affiliation{
 Department of Physics \& Astronomy, Amherst College, Amherst, Massachusetts 01002, USA
}
\author{J.F. Lin}
\affiliation{Department of Earth and Planetary Sciences, Jackson School of Geosciences, University of Texas at Austin, Austin, Texas 78712, USA}
\author{L.R. Hunter}
\affiliation{
 Department of Physics \& Astronomy, Amherst College, Amherst, Massachusetts 01002, USA
}

\date{\today}

\begin{abstract}

Terrestrial experiments that use electrons in Earth as a spin-polarized source have been demonstrated to provide strong bounds on exotic long-range spin-spin and spin-velocity interactions. These bounds constrain the coupling strength of many proposed ultralight bosonic dark-matter candidates. Recently, it was pointed out that a monopole-dipole coupling between the Sun and the spin-polarized electrons of Earth would result in a modification of the precession of the perihelion of Earth. Using an estimate for the net spin-polarization of Earth and experimental bounds on Earth's perihelion precession, interesting constraints were placed on the magnitude of this monopole-dipole coupling.  Here we investigate the spin associated with Earth's electrons.  We find that there are about $6 \times 10^{41}$ spin-polarized electrons in the mantle and crust of Earth oriented anti-parallel to their local magnetic field. However, when integrated over any spherically-symmetric Earth model, we find that the vector sum of these spins is zero. In order to establish a lower bound on the magnitude of the net spin along Earth's rotation axis we have investigated three of the largest breakdowns of Earth's spherical symmetry: the large low shear-velocity provinces of the mantle, the crustal composition, and the oblate spheroid of Earth.  From these investigations we conclude that there are at least $5 \times 10^{38}$ spin-polarized electrons aligned anti-parallel to Earth's rotation axis.  This analysis suggests that the bounds on the monopole-dipole coupling that were extracted from Earth's perihelion precession need to be relaxed by a factor of about 2000.
\end{abstract}

\maketitle

\section{\label{sec:motivation}Introduction}

Many extensions of the Standard Model suggest new bosonic particles, several of which (e.g. the axion) are also interesting dark-matter candidates \cite{safronova2018}. The virtual exchange of these bosons can result in new exotic (non-Standard Model) forces between particles \cite{Moody1984,dobrescu2006,fadeev2019}.  Searches for exotic spin-dependent forces provide important constraints on the coupling strength of these bosons \cite{clayburn2023,yan2015,zhang2023,romalis2020,romalis2022} as well as on torsion gravity \cite{hammond2002}. A comprehensive review of existing experimental and observational constraints on exotic spin-dependent interactions can be found in Ref.\ \cite{cong2024spindependentexoticinteractions}.  Spin-polarized electrons within Earth’s mantle and crust (geoelectrons) have been demonstrated to be a vast and valuable spin source in searches for exotic long-range spin-spin \cite{hunter2013} and velocity-dependent spin-spin \cite{hunterAng2014} interactions.  
Recently, it has been suggested that this large reservoir of spin-polarized geoelectrons can be combined with experimental measurements of the precession of the perihelion of Earth in order to establish bounds on exotic monopole-dipole couplings between the Sun and the spin-polarized electrons of Earth \cite{poddar2023}.  To establish such a bound, the magnitude and direction of the net electronic spin of Earth must be known.  Here we use the electron-spin model developed in Ref.\ \cite{hunter2013} to estimate the magnitude and direction of the net spin associated with these geoelectrons.  Assuming only that Earth’s mantle and crust are non-conductive and that their temperature, density, pressure, and chemical composition are spherically symmetric, we come to the general and somewhat surprising conclusion that the net spin of Earth's geoelectrons is zero. This is despite the fact that we estimate that there are about $6\times10^{41}$ spin-polarized geoelectron’s in Earth’s mantle and crust pointing anti-parallel to their local magnetic field. In order to determine a lower bound on the magnitude of the net spin along Earth's rotation axis we consider the breakdown of Earth's spherical symmetry associated with the large low shear-velocity provinces (LLSVPs), the crustal composition, and the oblate spheroid of Earth. We conservatively estimate that there are at least $5 \times 10^{38}$ spin-polarized electrons aligned anti-parallel to Earth's rotation axis. This analysis suggests that the bounds on the monopole-dipole coupling established in \cite{poddar2023} need to be relaxed by a factor of $\sim$ 2000.

\section{Methods}\label{sec:Setup}
\paragraph{Earth's Chemistry}Earth’s mass is distributed with $\sim$32 percent in the core, $\sim$68 percent in the mantle and $<$1 percent in the crust \cite{monroe2006physical}.  Electrical currents in the core are believed to be responsible for generating most of Earth's magnetic field.  Though the core is primarily composed of an Fe-Ni alloy, ab initio calculations suggest that, at core pressures and temperatures, it is energetically unfavorable for these metals to retain any unpaired electron spins.  This implies there is no net electron spin within the core \cite{alfe1998}.

In the presence of the geomagnetic field, some of the electrons in paramagnetic minerals within Earth's mantle and crust acquire a small spin polarization. Determining the magnitude and direction of this spin polarization requires knowledge of the temperature, the density and spin state of the unpaired electrons, and the magnetic field throughout the region. The unpaired electron-spin density and spin state in candidate host minerals can be calculated using results from deep-Earth mineral physics, geochemistry, and seismology \cite{irifune2010,hunter2013}.

The most abundant transition metal in the oxides and silicates of Earth's mantle, Fe, has a partially filled d-shell that dominates the resulting paramagnetism of the mantle minerals and crustal rocks. Other major rock-forming elements have closed electron shells with negligible contributions to the polarized spin density. The relevant physical parameters of the mantle oxides and silicates, and expected pressure-temperature profiles of the crust and mantle (geotherm) are used in the same manner as in Ref.\ \cite{hunter2013} to calculate the densities of the major mantle minerals at different depths \cite{poirier2000,Stacey_Davis_2008,lay2008}. These calculations agree well with the seismic preliminary reference Earth model (PREM) \cite{PREM_1981}, a standard model for the structures of Earth which determines the elastic parameters and densities throughout Earth using observation of seismic waves and other constraints. Most of the iron in Earth's crust and mantle is bound up in mineral lattices with either ferrous (Fe\textsuperscript{2+}) or ferric (Fe\textsuperscript{3+}) valence states. The Fe\textsuperscript{2+} high-spin (HS) state has four unpaired d-shell electrons and a total spin S = 2. The Fe\textsuperscript{3+} HS state has five unpaired d-shell electrons and a total spin S=5/2. The low spin (LS) state of Fe\textsuperscript{2+} and Fe\textsuperscript{3+} have total spin S = 0 and 1/2, respectively.  The relative densities of the specific spin states is determined as outlined in \cite{hunter2013}.  

The majority of the mantle's unpaired electrons exist in the Fe\textsuperscript{2+} HS state. When a spin-1/2 electron in this state interacts with the local geomagnetic field $\vb{B}$ on average it results in a polarized spin density anti-parallel to $\vb{B}$:  
\begin{equation} \label{eq:symmetry}
    \vb{<S_B>}=-N\frac{ \mu_B \vb{B}}{ k_B T}.
\end{equation}
Here $N$ is the number of unpaired electrons per unit volume in the S = 2 spin state, $\mu_B$ is the Bohr magneton, $T$ is the temperature, and $k_B$ is the Boltzmann constant.  Similar expressions for the unpaired electrons in the S=5/2, and S=1/2 are also proportional to $\vb{B}$ and differ only by a numerical factor \cite{hunter2013}. 

Note that the total number of spin-polarized electrons per unit volume only depends on the magnetic field $\vb{B}$, the number of unpaired electrons per unit volume $N$, the spin state of the mineral lattice S, and temperature $T$. The relevant geophysical models \cite{poirier2000,Stacey_Davis_2008,lay2008,PREM_1981} assume that Earth's  interior temperature, density, pressure and chemical composition depend only on the distance $r$ from Earth's center. Hence, Eq.\eqref{eq:symmetry} implies that over any spherical shell of radius $r$, the polarized spin density is simply proportional to the local magnetic field. Thus, if we demonstrate that the magnetic field vectors sum to zero for a spherical shell at a radius $r$, then we will also show that the net electron spin polarization at that radius also sums to zero.  Summing over all Earth radii outside of the core will then demonstrate that the net spin of Earth is in fact zero.  

\paragraph{Earth's Magnetic Field}The model we use to describe Earth's magnetic field is essentially the same as was used in Ref.\ \cite{hunter2013}.  It is based upon the World Magnetic Model (WMM) which is used by the U.S. Department of Defense, the U.K. Ministry of Defence, the North Atlantic Treaty Organization (NATO), and the International Hydrographic Organization (IHO)  \cite{nerc527808}. Our model assumes that the regions considered do not contain significant free currents and hence $\curl{\vb{B}}=0$.  In these regions, the magnetic field can be derived as the negative gradient of a scalar magnetic potential, $V_B$, which satisfies Laplace's equation.  In the WMM model, the coefficients of a 12th order associated Legendre solution for $V_B$ are chosen to yield the best fit to surface and satellite magnetic data. The coefficients for the expansion are produced jointly by the National Centers for Environmental Information (NCEI) and the British Geological Survey (BGS), and are usually updated every five years \cite{WMM_Coef}. The model is typically used for the magnetic field at and above the surface of Earth. However, since Earth's mantle and crust are primarily composed of insulating materials and have minimal free currents, the solutions can be extended to the mantle and crust with reasonable accuracy \cite{earthcore}. For any point outside the core radius, $R_C$, the WMM potential can be written as:
\begin{equation}\label{eq:potential}
    V_B(r,\theta,\varphi) =
    R_E \sum_{n=1}^N \sum_{m=0}^n \left(\frac{R_E}{r}\right)^{n+1} 
    [ g_n^m\cos{(m\varphi)}+h_n^m\sin{(m\varphi)} ] 
    P_n^m(\cos\theta),
\end{equation}
where $\vb{r}$ is the radial vector from the center of Earth to the point, $\theta$ is the point's co-latitude (the angle between $\vb{r}$ and Earth's rotation axis), $\varphi$ is the point's azimuthal angle, $R_E$ is Earth's radius, $P_n^m(\cos{\theta})$ is the $n$th degree, $m$th order associated Legendre function, and $g_n^m$ and $h_n^m$ are the coefficients given by the WMM. To extract the local magnetic field from this potential, one uses the equation 
\begin{equation}\label{eq:gradpotential}
\begin{aligned}
    \vb{B}(r,\theta,\varphi) = -\grad{V_B(r,\theta,\varphi)}.
\end{aligned}
\end{equation}

\begin{figure}[htp]
    \centering
    \includegraphics[width=10cm]{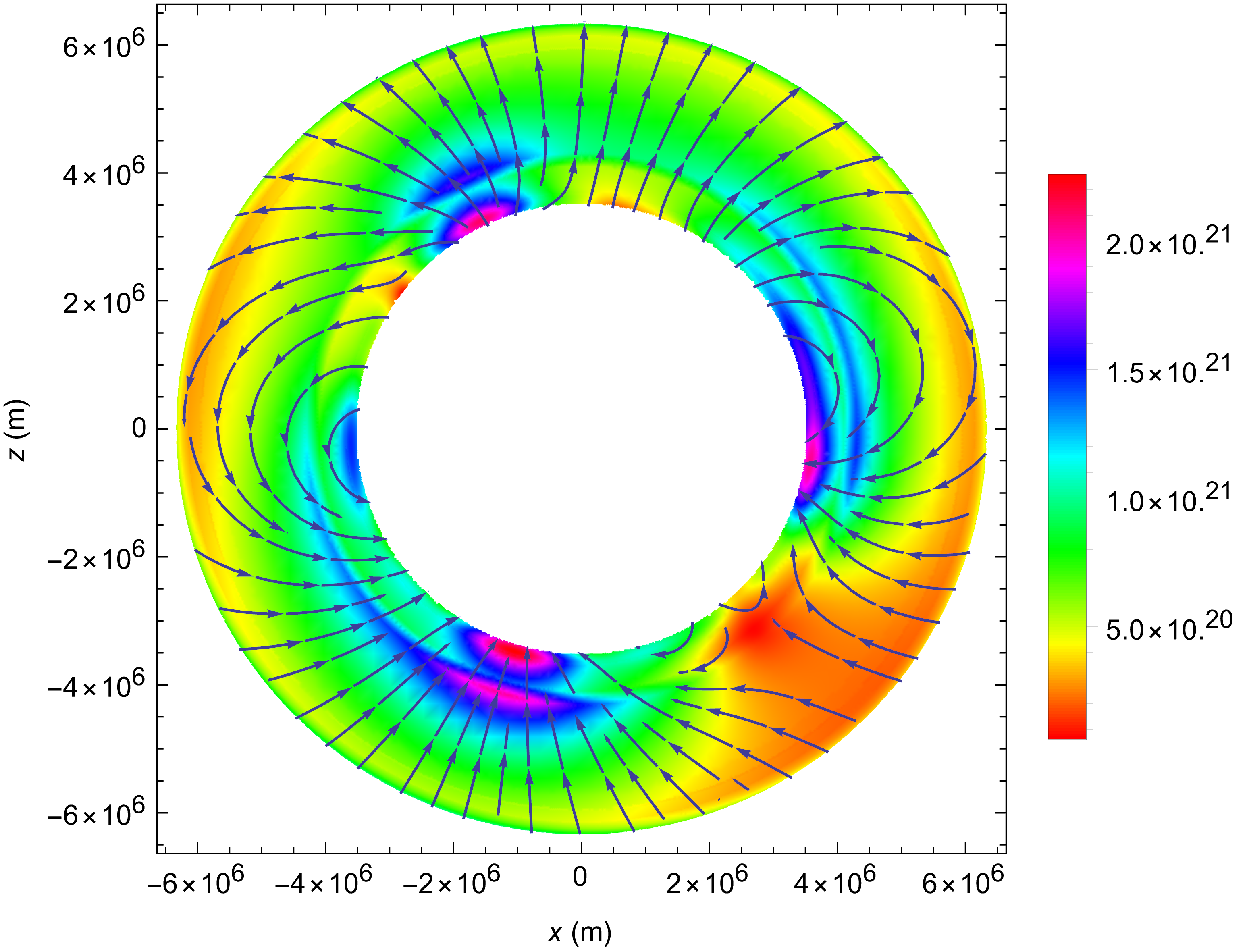 }
    \caption{The polarized electron-spin density on a plane that contains Earth's rotation axis (\textit{z}) and the prime meridian. The arrows indicate the local direction of the electron spins and the colored shading of the plot indicates the magnitude of the polarized electron-spin density in units of electrons per cubic meter. Only the spin components in the plane of the cross-sectional slice are shown. The electron-spin density within the core (white central circle) is assumed to be zero.}
    \label{fig:EarthSpin}
\end{figure}

\section{Calculation}

Following the procedures outlined in Ref.\ \cite{hunter2013}, we derived numerically an electron-spin map of Earth. Figure \ref{fig:EarthSpin} illustrates the complexity of the electron-spin distribution within Earth and suggests significant cancellation in the vector sum of Earth's oriented electron-spins. We performed numerical integrations of the electron-spins over Earth's volume using Mathematica and found a surprisingly high degree of cancellation of the net spin. This motivated us to investigate the problem analytically. Here we prove that within the WMM model assumptions there is an exact cancellation of the net spin of Earth. 

We begin by considering the direct integration of the $z$ component of the magnetic field over a spherical shell at a fixed radius $r > R_C$:
\begin{equation}
    \ {B_{net,z}} = \int_{} \vb{B} \cdot \Hat{\vb{z}} \dd{a}.
\end{equation}
We replace $\vb{B}$ with $-\grad{V_B}$ and insert the WMM potential (Eq. \eqref{eq:potential}):
\begin{gather}
    \ B_{net,z}= -\int_{} \grad{V_B} \cdot \Hat{\vb{z}}\dd{a}= 
\nonumber \\ -\sum_{n=1}^N \sum_{m=0}^n \int_{}\grad{\left(R_E\left(\frac{R_E}{r}\right)^{n+1} 
    [ g_n^m\cos{(m\varphi)}+h_n^m\sin{(m\varphi)} ] 
    P_n^m(\cos\theta)\right)}\cdot \Hat{\vb{z}}\dd{a}.
\end{gather}
All of our Legendre functions are functions of $\cos{\theta}$, so we omit the argument for simplicity in the following expressions. Writing out the gradient in spherical coordinates we find:
\begin{equation}
\begin{aligned}
    \vb{B}(r,\theta,\varphi) = -\grad{V_B(r,\theta,\varphi)} \
    =&\ \sum_{n=1}^N \sum_{m=0}^n (n+1)\left( \frac{R_E}{r}\right)^{n+2}
    [g_n^m\cos{(m\varphi)}+h_n^m\sin{(m\varphi)}]P_n^m \vu{r} \\
    &- \left(\frac{R_E}{r}\right)^{n+2}
    [g_n^m\cos{(m\varphi)}+h_n^m\sin{(m\varphi)}]
    \dv{P_n^m}{\theta} \vu*{\theta} \\
    &- \frac{m}{\sin\theta}\left( \frac{R_E}{r}\right)^{n+2}
    [-g_n^m\sin{(m\varphi)}+h_n^m\cos{(m\varphi)}]P_n^m \vu*{\varphi}.
\end{aligned}
\end{equation}
Substituting in $\vu{z} = \cos{\theta}\,\vu*{r} - \sin{\theta}\,\vu*{\theta}$ and evaluating the dot products yields:
\begin{equation}\label{A4}
   B_{z} = 
   \vb{B}(r,\theta,\varphi)\cdot\vu{z} = \sum_{n=1}^N \sum_{m=0}^n
   \left( \frac{R_E}{r} \right)^{n+2}
   [(n+1)\cos{\theta} P_n^m + \sin{\theta}\dv{\theta} P_n^m ]
   [g_n^m\cos{(m\varphi)} + h_n^m\sin{(m\varphi)}].
\end{equation}
We integrate each term of the associated Legendre expansion separately:
\begin{equation}
\begin{aligned} 
  \left( \frac{R_E}{r}\right)^{n+2}\int_0^\pi\int_0^{2\pi}   ((n+1)\cos\theta P_n^m + \sin\theta \dfrac{d}{d\theta}P_n^m)[g_n^m\cos{(m\varphi)}+h_n^m\sin{(m\varphi)}]r^2\sin{\theta}
     \dd{\varphi}\dd{\theta}
     \\ =  r^2\left( \frac{R_E}{r}\right)^{n+2}\int_{0}^{\pi}\sin\theta[(n+1)\cos\theta P_n^m + \sin\theta\dfrac{d}{d\theta}{P_n^m}]   
     \int_{0}^{2\pi}[g_n^m\cos{(m\varphi)}+h_n^m\sin{(m\varphi)}]\dd{\varphi} \dd{\theta}.
\end{aligned}     
\end{equation}
Examining first the $\varphi$ integral we have
\begin{equation}
    \int_{0}^{2\pi}[g_n^m\cos{(m\varphi)}+h_n^m\sin{(m\varphi)}]\dd{\varphi}.
\end{equation}
 When \(m\geq 1\) the integral is zero. For \(m=0\), the expression for the integral reduces to:

\begin{equation}
     2\pi g_n^0 r^2\left( \frac{R_E}{r}\right)^{n+2}\int_{0}^{\pi}((n+1)\cos\theta P_n^0 + \\ \sin\theta\dfrac{d}{d\theta}{P_n^0})\sin{\theta}\dd{\theta}. 
\end{equation}
Now only looking at the theta integral, we have
\begin{equation} \label{A11}
    (n+1)\int_{0}^{\pi} \cos \theta P_n^0 \sin\theta \dd{\theta} +\int_0^\pi \sin \theta \dfrac{d}{d\theta}{P_n^0} \sin\theta \dd{\theta}.
\end{equation}
The WMM model includes a factor of $(-1)^m$ in the definition of the associated Legendre functions.  Including this factor, we find that \(P_1^0=\cos\theta\), \(P_1^1=-\sin\theta\) and $\dfrac{d}{d\theta}P_n^0=P_n^1$ \cite{Derivatives}. Substituting these into Eq. \eqref{A11} yields 
\begin{equation}\label{A10}
    (n+1)\int_{0}^{\pi}P_1^0 P_n^0 \sin\theta \dd{\theta} -\int_{0}^{\pi} P_{1}^{1}P_n^1 \sin\theta \dd{\theta}.
\end{equation}
We use the orthogonality relation (Eq. 10.8 of Ref.\ \cite{Boas:913305}), 
\begin{align}
    \int_0^\pi P_n^m P_l^m \sin\theta d\theta = \frac{2(n+m)!}{(2n+1)(n-m)!}\delta_{l,n}, \label{eq:ortho}
\end{align}
to evaluate each integral in Eq.\eqref{A10}. The integral is zero when $n\neq 1$. For \(n\) = 1 the first integral has a value of 2/3 and the second has a value of 4/3. Including the $n+1 = 2$ factor, the two \(n\) = 1 terms cancel. Thus, when integrated over theta and phi, all of the terms in Eq.\eqref{A4} sum to zero and hence the integration of $B_z$ over any spherical shell with radius $r>R_C$ reduces to zero. It is interesting to note that this proof indicates that the order of the Legendre expansion and the values of the coefficients $ g_n^m$ and $ h_n^m$ are irrelevant to the conclusion that the integral of $\vb{B}$ over any spherical shell is identically zero.

One can argue that since the vanishing of the angular integration of $B_z$ does not depend on the values of $g_n^m$, $h_n^m$ or $r$, the integrated magnetic field component along any direction must sum to zero. Recall that in the WMM model, the coefficients $g_n^m$ and $h_n^m$ are chosen to reproduce Earth's magnetic field when $\Hat{\vb{z}}$ lies along Earth's rotation axis.  Since the associated Legendre functions form a complete set of solutions to Laplace's equation, we could just as easily have chosen our $\Hat{\vb{z}}$ axis to be in Earth's equatorial plane.  The only thing that would change in the solution is that we would require different values of $g_n^m$ and $h_n^m$ to match Earth's magnetic field.  Since our conclusion that the net $\Hat{\vb{z}}$ component of $\vb{B}$ must vanish is independent of the values of $g_n^m$ and $h_n^m$, we infer that the net component of $\vb{B}$ in the equatorial plane (or for that matter any other direction) must also be zero.  Hence, we arrive at the important conclusion that the integral of $\vb{B}$ over any spherical shell with radius $r$ greater than the core radius, $R_C$, must be zero.

In appendix \ref{app:app1} we present an alternative proof of the complete cancellation of the net magnetic field over a spherical shell. In this approach, we apply the results of a classical magnetostatics problem to show that the magnetic field integrated over any spherical shell outside of a region of constant currents must sum to zero. This approach is independent of the representation used for the magnetic field.

\section{Discussion and Conclusion}
Here we have demonstrated that using a standard spherically-symmetric geophysical model of Earth and its properties, the net electron spin associated with Earth is zero. The $z$-component of the net electron spin plays a critical role in creating Earth's anomalous perihelion precession as the equatorial components cancel daily due to Earth's rotation.  Though it is not explicitly stated in Ref.\ \cite{poddar2023}, it appears from their analysis that the authors have assumed that there are $10^{42}$ electrons spin polarized along $\Hat{\vb{z}}$. That number is an order of magnitude estimate the authors inferred from Ref.\ \cite{hunter2013}. However, Ref.\ \cite{hunter2013} does not estimate the number of electrons oriented along $\Hat{\vb{z}}$, but is rather an order of magnitude estimate of the number of electrons in the mantle and crust that are oriented anti-parallel to their own local magnetic fields. We have run a numerical integration using the model described here and find that there are about $6 \times 10^{41}$ spin-polarized electrons within the mantle and crust with their spins anti-parallel to their local magnetic field, consistent with the earlier order of magnitude estimate.  However, as we have demonstrated analytically here, this model does not yield any net spin along $\Hat{\vb{z}}$.  

If one treats Earth as a point-like electron-spin source, then to establish an upper bound on the monopole-dipole coupling from the measurements of its perihelion precession, one needs to establish a minimum net electron spin along $\Hat{\vb{z}}$.  We are unaware of any previous lower bound on this quantity. One might hope to establish a lower bound on the net electron spin along $\Hat{\vb{z}}$ by considering the limitations of some of the approximations that are made in the present analysis.  
We note that at any point the electron-spin along the z direction is proportional to $B_z$.  The dominant term in Earth's magnetic field is dipolar and is described by the equation (page 255 of Ref.\ \cite{grifem}) 
\begin{equation}\vb{B}_{dip} = \frac{\mu_0 m_z}{4\pi r^3}(2\cos{\theta}\,\Hat{\vb{r}} + \sin{\theta}\,\Hat{\vb{\theta}}) 
\end{equation} where $m_z$ is the z component of Earth's magnetic dipole moment.  Recalling that $\vu{z} = \cos{\theta}\,\vu*{r} - \sin{\theta}\,\vu*{\theta}$ we perform the dot product $\vb{B}_{dip}\cdot \Hat{\vb{z}}$ to project out the z component of the dipole field, \begin{equation} B_{dip,z} = \frac{\mu_0m_z}{4\pi r^3} (2\cos^2{\theta} - \sin^2{\theta})=\frac{\mu_0m_z}{4\pi r^3} (3\cos^2{\theta} - 1)=2\frac{\mu_0m_z}{4\pi r^3}P_2(\cos{\theta}).
\end{equation}
Due to the orthogonality of the Legendre polynomials, When $B_z$ is multiplied by the density of unpaired electrons and integrated over Earth's volume, densities that have a significant $P_2(\cos{\theta})$ component will usually dominate the integration.

The most obvious breakdown of the assumption of spherical symmetry is that Earth is an oblate spheroid with its polar radius about 21 km less (about 0.3\%) than its equatorial radius \cite{Snyder_1987}.  The dominant correction to the electron-spin along $z$ is associated with the quadrupole term, $P_2(\cos\theta)$, implied by this flattening of Earth.  We undertake such a correction in Appendix \ref{app:app2} and estimate that this departure from spherical symmetry results in $\sim 3.2 \times 10^{38}$ electrons spin-polarized along $-\hat{\bf{z}}$.

Another prominent non-radial heterogeneity of Earth is associated with the different compositions of different layers of Earth's crust and uppermost mantle. The crust can be further divided into layers of ice, water, sediments and crystalline crust. The mass and iron densities of these different layers as well as their radial and angular distributions can potentially lead to a net electron spin along $\hat{\bf{z}}$. At shallow depths, the largest heterogeneity is between the upper continental crust and Earth's oceans which have very low iron densities \cite{Huheey1993}. Below the oceans, the thin basalt-based oceanic crust has a higher iron density than its primarily granite continental counterpart at the same depth. At depths below the oceanic crust, the dominant non-radial heterogeneity is between the continental crust and the uppermost mantle.  In Appendix \ref{app:app4} we use a global crustal model and estimates of the iron fractions of the various layers to estimate their likely contribution to the net electron spin along $\hat{\bf{z}}$. We find that these crustal heterogeneities yield $(2.2 \pm 1.5) \times 10^{38}$ electrons spin-polarized along $-\hat{\bf{z}}$. 

Similarly, one might suspect that non-radial heterogeneities in the mantle might to some degree spoil its electron-spin cancellation. In the mantle the largest known non-spherical structures detected by seismic observations are two LLSVPs that have been identified in seismic studies \cite{Garnero2016}. It has been speculated that these regions might have higher iron concentrations, mass densities and/or temperatures than their surrounding mantle materials. In Appendix \ref{app:app3} we conservatively estimate that these LLSVPs contribute at least $\sim 1.6 \times 10^{38}$ electrons spin-polarized along $-\hat{\bf{z}}$.

If one hopes to take advantage of the abundant electron-spin polarization in Earth's mantle and crust to bound exotic spin interactions, it may be necessary to go beyond the model of Earth as a point-like spin source. One might instead use the radial fall off of the exotic spin-dependent potential over a finite sized Earth to spoil the cancellation observed here.  This is precisely what is done to achieve the bounds established by terrestrial experiments \cite{hunter2013,hunterAng2014}.  Unfortunately, for an interaction between Earth's geoelectrons and the Sun, the sensitivity achieved by such a model will be suppressed by factors of order $D/R_E$, where $R_E = 6.4 \times 10^3$ km is the Earth radius and $D = 1.5 \times 10^8$ km is the distance to the Sun. This suppression is likely greater than $10^4$.  Note that no similar large suppression occurs for related experiments that take place on Earth's surface. 

Given the multiplicity of possible sources that can spoil the assumption of a spherically symmetric Earth and their sensitivity to the parameters and models used, it is difficult to place exact lower bounds on the net electron spin along $\Hat{\vb{z}}$.  Here we have identified and conservatively estimated what are likely to be the most important of these sources.  All three of these sources yield electron-spin polarizations along $\Hat{\vb{z}}$ that have the same sign and similar magnitudes.  To arrive at a conservative lower bound for the number of electrons spin-polarized along $-\Hat{\vb{z}}$ we sum the lower bounds associated with each of these heterogeneities.  Including the model uncertainties, we arrive at a conservative estimate that the number of electrons spin-polarized along $-\Hat{\vb{z}}$ is likely greater than $\sim 5 \times 10^{38}$. Further refinement of this bound will require more precise determinations of Earth's elemental abundances and additional self-consistent analysis with greater detail.  A future calculation should also incorporate more recent information on the spin transitions in ferropericlase \cite{FU_EPSL_2018} and bridgmanite \cite{Lin2016, Fu_GRL_2018} as well as on the iron partitioning in ferropericlase \cite{Xu2017}.

Our lower bound on the net electron spin along $\hat{\bf{z}}$ is 2000 times lower than the value assumed in \cite{poddar2023}.  This suggests that their published bounds on the monopole-dipole coupling determined from Earth's perihelion precession need to be relaxed by this same factor.

\section{ACKNOWLEDGEMENTS}
This work was supported by National Science Foundation Grant No. PHY-2110523.  The authors thank Prof. William Loinaz and Prof. Kannan Jagannathan for useful conversations pertaining to the proofs presented here. This work was performed in part using the high-performance computing equipment at Amherst College obtained under National Science Foundation Grant No. 2117377.

\appendix 

\section{Representation-Independent Derivation}\label{app:app1}

We begin this alternative proof with Eq. 5.93 of Ref.\ \cite{grifem}.  The equation implies that the average magnetic field, over a sphere of radius $R$, due to steady currents inside the sphere is,
\begin{gather}
    \vb{B}_\text{av} =
    \frac{\mu_0}{4\pi} \frac{2\vb{m}}{R^3},
\end{gather}
where $\vb{m}$ is the total dipole moment generated by currents within the sphere. The integral of $\vb{B}$ over the volume of the sphere is therefore
\begin{gather}\label{grif}
    \int_V \vb{B} \dd{\tau} =
    \vb{B}_\text{av} \frac{4\pi}{3} R^3 =
    \frac{2 \mu_0}{3} \vb{m}.
\end{gather}
\\

Now consider a thin spherical shell in Earth's mantle or crust between radii $r$ and $r+\dd{r}$, where $R_C<r<R_E$. Let the radius $r$ correspond to spherical volume $V_1$, and the radius $r+\dd{r}$ correspond to the spherical volume $V_2$. Assume all magnetic field sources are in the core, which is contained in $V_1$. Therefore the net magnetic field in the thin spherical shell is: 
\begin{gather}
        \int_{V_2} \vb{B} \dd{\tau} - \int_{V_1} \vb{B} \dd{\tau} =
    \frac{2 \mu_0}{3} \vb{m} - \frac{2 \mu_0}{3} \vb{m} = 0.
\end{gather}

This implies that the integral of the magnetic field over any sphere of radius $r$, where $R_C<r<R_E$, is zero. The electronic spin at any point in the mantle and crust is a function of $\vb{B}$, temperature, and unpaired electron density (Eq. \eqref{eq:symmetry}). For any given radius, the temperature and unpaired electron density are constant \cite{hunter2013}, and the net $\vb{B}$ is zero, so for any spherical shell in the mantle and crust the net electronic spin over the spherical shell is zero. Therefore, when summed over the entire mantle and crust, the net electronic spin is also zero. 

\section{Quadrupolar Mass Contribution Estimate}\label{app:app2}

Here we approximate the contribution of Earth's quadrupolar mass distribution to the net electron spin along $\Hat{\bf{z}}$. We use the hydro-static model of Nakiboglu \cite{NAKIBOGLU1982302}  to create a first-order correction to our spherically symmetric electron-spin density, $\rho_0(r)$.  In this model 
\begin{equation}
\rho(r,\theta)=f_2(r)\rho_0(r) P_2(\cos{\theta}) , 
\end{equation}
where $f_2(r)$ is a negative dimensionless zonal harmonic coefficient that accounts for the variation in the ellipsoidal flattening factor as $r$ decreases.  In \cite{NAKIBOGLU1982302} the value of $f_2(r)$ has been adjusted to yield a best fit to the moments of inertia of Earth, the gravitational potential outside of Earth, and Earth's surface flattening while maintaining Earth's mass, the radial density of the PREM model and hydro-static equilibrium. To facilitate integration, we have fit the values of $f_2(r)$ contained in Table \ref{tab:table1} of \cite{NAKIBOGLU1982302} over the radii contained in the mantle and crust to a 4th order polynomial.  Numerically integrating this spin density over the mantle and crust yields a net polarization of $\sim 3.2 \times 10^{38}$ electrons spin-polarized along $-\Hat{\bf{z}}$. In order to get a sense of the model dependence of this analysis, we have also evaluated it for the extreme case of a uniform-density Earth where $f_2(r)$ is assumed to be a constant equal to its value at Earth's surface, $-0.00223$ \cite{NAKIBOGLU1982302}. This analysis yields a net polarization of $\sim 3.6 \times 10^{38}$ electrons spin-polarized along $-\Hat{\bf{z}}$, suggesting that the result is fairly robust to the precise internal model used. We assume an uncertainty in our central value of about 10\%. We note that the sign of the net electron spin is consistent with Fig. \ref{fig:EarthSpin}, which generally has more negative electron-spin components in the equatorial plane where there is more mass, and more positive components in the direction of the poles where there is less mass.

\section{Crustal Contribution Estimate}\label{app:app4}

Here we model the effect of non-radial heterogeneity in the composition of Earth's crust and uppermost mantle. These inhomogeneities can be described using the ECM1 global crustal model \cite{MOONEY2023104493}. This model identifies Earth properties in a one-degree grid including the thickness of water, ice, sediment, and crustal layers. The model is based on 19,200 seismic measurements and provides a three-dimensional density distribution of the crust and uppermost mantle. Moreover the ECM1 assigns one of 12 crustal types to each geographically-defined tile. These crustal types can be further categorized as either continental or oceanic allowing assignment of the iron fractions of Table 1 to these regions. To estimate the effects on the net electron-spin polarization along $\hat{\bf{z}}$ we combine the ECM1 model with these iron fraction estimates. Integrating over shells of varying depths and thicknesses we are able to determine the contributions of different layers to the net electron-spin along $\hat{\bf{z}}$. This result depends highly on the assumed iron fractions and mass density distributions. To quantify the uncertainties associated with the allowed iron fractions we select the iron fractions in Table 1 that maximize or minimize the number of electrons spin-polarized along $\hat{\bf{z}}$. We integrate over a volume which extends 40 km below and 10 km above sea level and includes the majority of the crust's heterogeneity. This integration yields $(2.2 \pm 1.5) \times 10^{38}$ electrons spin-polarized along $-\hat{\bf{z}}$. To explore the effect of different mass distributions we repeat the integration using the densities and depths given by CRUST 1.0 \cite{2013EGUGA}, an alternative global crustal model, and find similar results.\\

\begin{table*}[ht]
\caption{\label{tab:table1}Weight percents of the total iron expressed as the calculated ferrous amount (FeOt) for the crust and uppermost mantle layers. As in the ECM1 model, we divide the oceanic and continental crust into its constituent ice/water, crustal, and sedimentary layers. Values for the upper mantle are also shown. The middle row for each gives the literature recommended value. To estimate the model dependence of these parameters (see text), 1-$\sigma$ variations in these central values are considered in the rows marked min and max. For the lower continental crust, 1-$\sigma$ variations are unavailable and instead a range of published concentrations are given. We are unable to identify uncertainty estimates for the thin sedimentary layers. {}\textsuperscript{\textdagger}Ref.\ \cite{Le_Maitre2005} is used to determine the FeOt weight percent from the FeO and Fe$_2$O$_3$ weight percents of Ref \cite{Yaroshevsky_2006}. In this model, we assume that all of the iron is in the HS state.$^a$\cite{HANSWEDEPOHL19951217}.$^b$\cite{Yaroshevsky_2006}.$^c$\cite{RudnickGao2014}.$^d$\cite{Huheey1993}.$^e$\cite{WHITE2014457}.$^f$\cite{Walter2014}.$^g$\cite{Taylor1985, Taylor1995}.}
\begin{ruledtabular}
\begin{tabular}{>{\centering\arraybackslash}p{1.8cm}|>{\centering\arraybackslash}p{2cm}>{\centering\arraybackslash}p{1.8cm}>{\centering\arraybackslash}p{1.8cm}>{\centering\arraybackslash}p{1.8cm}|>{\centering\arraybackslash}p{1.6cm}>{\centering\arraybackslash}p{1.9cm}|>{\centering\arraybackslash}>{\centering\arraybackslash}p{1.5cm}>{\centering\arraybackslash}p{1.5cm}} 
 &\multicolumn{4}{c|}{Continental}&\multicolumn{2}{c|}{Oceanic}&\multicolumn{2}{c}{Both}\\
 Considered Range &  Sedimentary &  Upper Continental &  Middle Continental &  Lower Continental &  Sedimentary & Oceanic Basaltic & Ice/Water & Upper Mantle \\ 
\hline
{Min FeOt}   & -    &4.51&   5.22&7.47$^a$&-&8.79&-&8.04\\
 Avg FeOt & 5.16{}\textsuperscript{\textdagger}$^b$    &5.04$^c$&   6.02$^c$&8.57$^c$&4.32{}\textsuperscript{\textdagger}$^b$&10.36$^e$&$\sim$0$^d$&8.27$^f$\\
 Max FeOt& -    &5.57&   6.82&10.6$^g$&-&11.93&-&8.41\\
\end{tabular}
\end{ruledtabular}
\end{table*}

\section{LLSVPs Contribution Estimate}\label{app:app3}

The LLSVPs are two broad regions in the lowermost mantle below Africa and the Pacific, characterized by percent-level perturbations of seismic wave speeds. Both anomalies extend laterally for thousands of kilometers and extend vertically $\sim$ 1,200 km from the core–mantle boundary. These perturbations reflect thermal, density, and/or compositional heterogeneity in those regions. A number of origin mechanisms have been proposed, including that the LLSVPs represent accumulations of subducted oceanic crust \cite{Christensen1994}, primordial thermochemical piles \cite{Lee2010}, a residue of basal magma crystallization \cite{Labrosse2007}, or remnants of a protoplanet theorized to have struck Earth \cite{Yuan2023}. These various origins often lead to different LLSVP material compositions, densities, and/or temperatures than those of the surrounding mantle material.

It has been suggested that these LLSVP regions may be enriched with as much as $\sim$40\%  more iron than the surrounding mantle at 2,800 km depths \cite{Deng2023}. Iron enrichment of the LLSVPs has also widely been suggested by previous studies \cite{Mosca2012,Trampert2004,DESCHAMPS2012,DESCHAMPS2019,Vilella2021}. The density anomalies associated with the LLSVPs remain actively debated. Tidal tomography-based studies found that the mean density of the lower two-thirds of the two LLSVPs is $\sim$ 0.5\% higher than that of the surrounding mantle. On the contrary, a study using Stoneley modes suggests a negative density anomaly within LLSVPs, without excluding the possibility of a high-density anomaly within the lowermost regions of the LLSVPs \cite{Koelemeijer2017}. A resolution to these discrepancies is presented in Ref.\ \cite{Deng2023} which models the 3D chemical composition and thermal state of the lower mantle based on seismic tomography and mineral elasticity data. Ref.\ \cite{Deng2023} concludes that the lowermost LLSVP regions are denser than the ambient mantle but at shallower depths the structures are less dense than the surrounding mantle. 

Acknowledging the challenges related to understanding the nature, origin, and morphology of the LLSVPs we attempt to estimate the minimum net electron spin along $\Hat{\bf{z}}$ that the LLSVPs might contribute. We use the work of Ref.\ \cite{Cottaar2016} to determine the spatial extent and location of the LLSVPs. That clustering analysis contrasts five global shear wave speed tomographic models and infers the morphology and volume of the LLSVPs. In our estimate we specifically use the majority ($m\ge3$) slow cluster vote map of Ref.\ \cite{Cottaar2016}. This represents a lower bound as it likely underestimates the volume of the LLSVPs as 6.7\% of the mantle. 

Within the LLSVPs we use the density and temperature variation inferred from the inversion of the GLAD-M25 model \cite{GladM25} described in Ref.\ \cite{Deng2023}. At depths below 2800 km we assume an iron fraction 40\% greater and a mass density 0.7\% greater than those assumed in Ref.\ \cite{hunter2013}. At depths shallower than 2700 km we assume a mass density 0.2\% less than Ref.\ \cite{hunter2013} while leaving the iron fraction unchanged. This reduction of the mass density is consistent with that suggested in Ref.\ \cite{Koelemeijer2017}. In the intermediate region between 2700 km and 2800 km we assume a mass density 0.25\% greater and an iron fraction 20\% greater. We also assume that the temperature in the LLSVPs is 20\% higher than the surrounding mantle material. From Eq. \eqref{eq:symmetry} this results in a 20\% reduction of all of the electron spins in these regions. Numerically integrating the contributions of these regions we find that they yield $\sim 4.1  \times 10^{38}$ electrons spin-polarized along $-\hat{\bf{z}}$. However, this result is highly model dependent. If the iron fractions of the model are reduced by a factor of 2 as suggested by the inversion of the SP12RTS model \cite{SP12RTA} described in Ref.\ \cite{Deng2023}, then the net electrons spin-polarized along $-\hat{\bf{z}}$ is reduced to $\sim 1.6  \times 10^{38}$. We regard this as a reasonable lower bound on the polarization along $-\hat{\bf{z}}$.  We note that the sign of the effect, based on Fig. \ref{fig:EarthSpin}, matches the expectation that one would have due to the additional polarizable spins the LLSVP models preferentially add to the equatorial regions.

\nocite{*}

\bibliography{apssamp}

\end{document}